\documentstyle[12pt,epsfig]{article}

\def\Title#1{\begin{center} {\Large {\bf #1} } \end{center}}

\begin{document}

\Title{HERA-B Status}

\bigskip\bigskip

%+\addtocontents{toc}{{\it A.B. Author}}
%+\label{authorStart}

\begin{raggedright}  

{\it M. Medinnis\index{Medinnis, M.}\\
DESY Zeuthen \\
Platanenallee 6 \\ 
D-15738 Zeuthen \\
 Germany }
\bigskip\bigskip
\end{raggedright}

% index item, please mark it.  \index{index item}

\section{Introduction}

HERA-B\cite{hbprop,hbtdr} is designed for studies of B-physics at
DESY's HERA proton-lepton storage ring in Hamburg, Germany. The
physics program calls for measuring production and decay properties of
B and B$_s$ mesons with emphasis on CP violation, particularly in the
B$^o$ $\rightarrow$ J/$\psi$ K$^o_s$ decay channel.

Collisions of protons in the 920 GeV beam of HERA with a fixed target
consisting of up to 8 wires surrounding the beam produce B-mesons
whose decay products are measured in the HERA-B spectrometer.  HERA-B
will operate with a 40 MHz collision rate which, given HERA's 10 MHz
bunch-crossing rate, implies an average of 4 interactions per bunch
crossing.  A sophisticated multi-level triggering system is needed to
reduce the overwhelming background from inelastic proton-nucleon
collisions to a rate suitable for transfer of data to mass storage.

As of this writing (Dec. $'$99), the experimental apparatus is nearing
completion and the triggering system is being commissioned. This paper
will briefly describe the experiment and summarize the current status
of the major detector components. A more detailed account may be found
in reference \cite{Padilla} and subsystem-specific papers in the same
volume.

\section{The detector}

\begin{figure}[htb]
\begin{center}
\epsfig{file=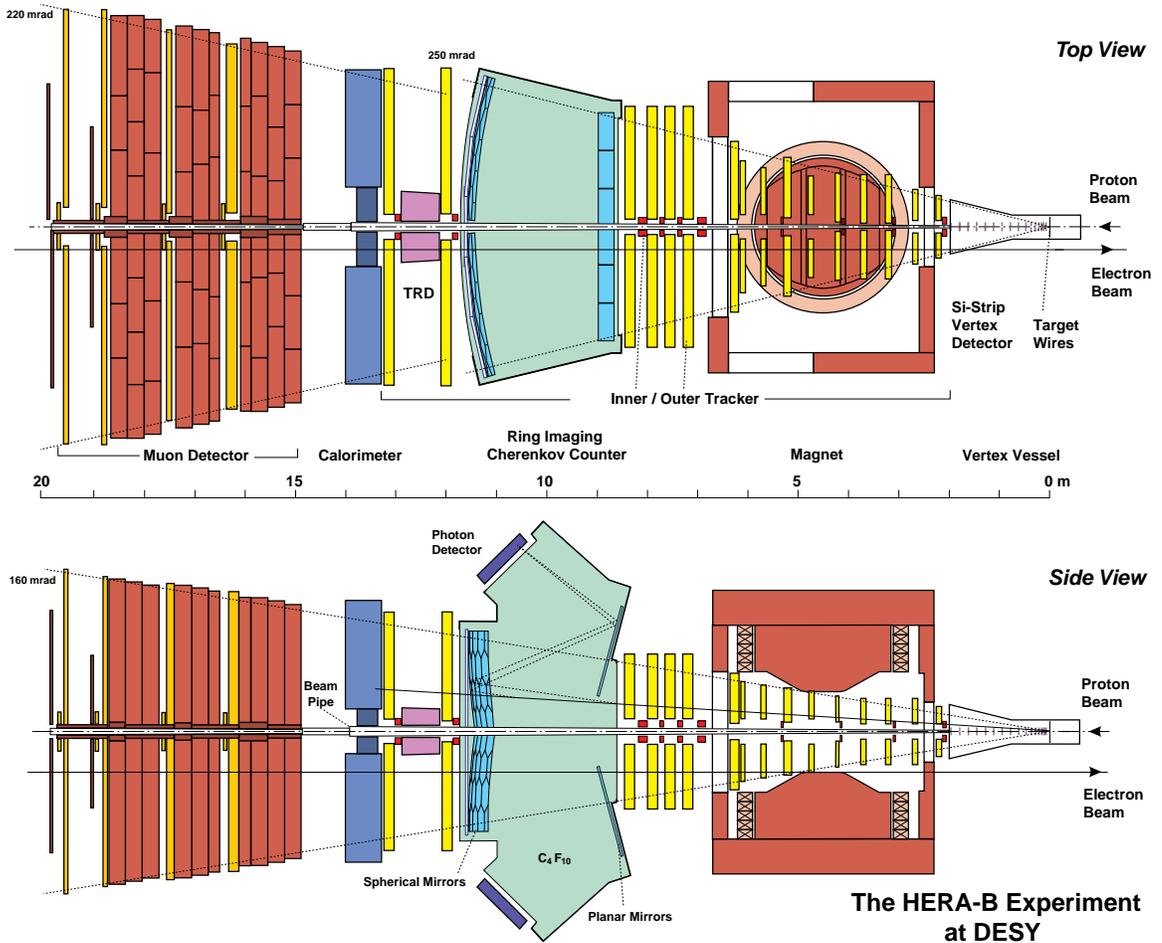,width=\textwidth}
\caption{The HERA-B spectrometer}
\label{fig:Spect}
\end{center}
\end{figure}

The apparatus is shown in Fig.~\ref{fig:Spect}. The proton beam enters
from the right on the sketch and first sees the target wires then
continues its journey through the spectrometer inside the
500\,$\mu$m thin aluminum beam pipe. Immediately downstream of the
target, one finds the eight super-layers of the vertex detector which
occupy the first two meters of the detector. A thin vacuum window is
located just before the last station. The main tracking system starts
immediately afterwards as does the large aperture main spectrometer
magnet. The super-layers of the tracker are divided into inner and
outer regions. The inner tracker covers the first 20\,cm from beam
center. The outer tracker takes over from there and extends the
coverage to 250\,mrad horizontally and 160\,mrad vertically.

Three of the tracking stations shown in the magnet are a combination
of cathode pad chambers (outer region) and gas pixel chambers (inner
region) which will be used in the trigger for identifying high-p$_t$
hadrons. A RICH counter occupies the region between 8.5\,m and
11.5\,m.  This is followed by another tracking layer, a transition
radiation detector which supplements electron identification in the
low-angle region, a tracking layer, then an electromagnetic
calorimeter. The calorimeter is followed by a muon detector which is
divided into three iron/concrete filters and 4 tracking stations. The
muon trackers are divided into inner and outer sections in the manner
of the main tracker.

\subsection{Vertex detector system}

\begin{figure}[htb]
\begin{center}
\epsfig{file=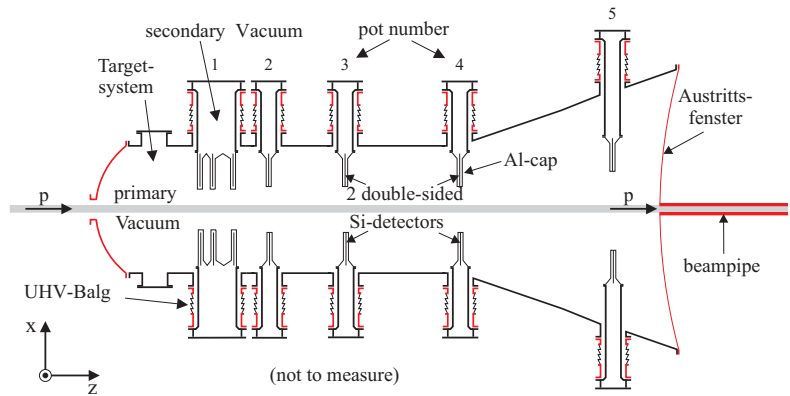,width=11cm,
  clip=, bbllx=98 ,bblly=362 ,bburx= 487,bbury= 576}
\caption{The vertex detector system}
\label{fig:vds}
\end{center}
\end{figure}

A horizontal section of the VDS is sketched in Fig.\ref{fig:vds}. (A
vertical section would look nearly the same.) Seven super-layers
\footnote{An eigth layer (not shown) is positioned just downstream of
  the exit window on an immovable mount.}  are each comprised of four
``quadrants'' each of which holds two double-sided silicon wafers. The
wafers are configured as 50 $\mu$m-pitch strip detectors with strips
oriented at $\pm2.5^o$ angles to the vertical or horizontal. The
detectors are mounted in thin movable aluminum caps (so-called ``Roman
pots'') which can be moved away from the beam to allow for filling.
The pots are evacuated but separated from the primary machine vacuum.
When in their final positions, the outer perimeter of the 4 detectors
of one super-layer describes a square centered on the beam with a
square hole to allow passage of the beam.

{\bf Status:} 85\% of the detector has been operating routinely since
summer of this year. The remainder is being installed in the present
shutdown. The detector is performing well with signal to noise ratios
of 20 or better. 

\subsection{Inner tracker}

The original design called for a ``classic'' micro-strip gas chamber
(MSGC) \cite{hbitr}.  Early tests with x-rays showed that the chambers
would withstand the anticipated dose rates for HERA-B but when tested
in a hadron beam, the chambers sustained considerable anode damage due
to sparking in a matter of hours. The problem was traced to heavy ions
produced when charged tracks traverse the cathode/anode wafer. After
some delay, a new design was found which uses a ``GEM'' foil (for gas
electron multiplier) positioned between the drift electrode and the
anode/cathode plane to provide additional gas amplification thus
lowering the needed amplification in the region of the anode.
Dose-related effects are still observed, nonetheless, the solution
fulfils our requirements.

{\bf Status:} Production and installation is underway with completion
scheduled for February of next year. Several chambers have been
installed, commissioned, and are operating on a regular basis. First
analysis of data indicate that the chambers are meeting
specifications. The readout electronics currently in use do not permit
their forseen use in the tracking phase of the first level trigger.
The readout of stations needed for the first level trigger will be
upgraded in time for the running period of 2001.

\subsection{Outer tracker}

The outer-tracker is built from honeycomb drift modules with hexagonal
cells of 5\,mm and 10\,mm.  The cathodes are made from a
carbon-impregnated resin ``Pokalon-C ''.  In situ tests of the
chambers in 1997 revealed that the original design would not withstand
the high radiation environment required for HERA-B operation and
forced considerable additional R\&D and substantial delays. Solutions
to the various aging problems have since been found.  The main
departures from the original design are that the Pokalon-C is now
gold-coated and a change of gas mixture was made, from from
Ar/CH$_4$/CF$_4$ to Ar/CO$_2$/CF$_4$.

{\bf Status:} All but one super-layer is now installed and operating
routinely.  Dead and noisy channel counts are of order 1\%.  The
missing super-layer is scheduled for installation in January.  The
alignment and calibration is in progress.

\subsection{RICH detector}

Cherenkov light produced in the 2.5\,m long C$_4$F$_{10}$ radiator gas
is focused by an array of spherical mirrors onto focal planes of lens
systems which then focus the light to an array of multi-anode
phototubes.

{\bf Status:} The detector is fully commissioned and in routine
operation. The average photon yield for $\beta$ = 1 rings is 35, as
expected.

\subsection{Electromagnetic calorimeter}

The calorimeter is of Shashlik design consisting of
cells of 11\,cm $\times$ 11\,cm transverse dimensions which are
stacked to form a wall some 6\,m in length and 4\,m in height. 
The calorimeter is sub-divided into 3 regions. The
cells of the innermost region use tungsten as a radiator and are
viewed by a 5 $\times$ 5 array of phototubes. Both middle and outer
sections have lead radiators. The cells of the middle region are
viewed by 4 phototubes, and those of the outer region are serviced by
a single phototube.

{\bf Status:} The calorimeter is fully installed and equipped with
phototubes. The inner and middle regions are also equipped with read
out electronics. The read out is scheduled for completion by end
January, 2000.  The calibration is continually being refined and
presently is about 7\%.

\subsection{Muon detetctor}

The muon detector consists of 4 superlayers. 
The first 40\,cm of each superlayer is covered by gas pixel chambers.
The outer regions of the first two superlayers are covered by 3 double
layers of tube chambers oriented at 0$^o$ and $\pm$20$^o$ with respect
to the vertical. The third and fourth superlayers are of similar design
but have, in addition, cathodes segmented into pads which are read out
separately, for use in triggering.

{\bf Status:} The system is fully installed and operational.
Occupancies are as expected and coincidence rates between the third
and fourth superlayers are also as expected.

\section{Trigger and data acquisition}

\subsection{First level trigger}

The task of the first level trigger (FLT) is to reduce the 10 MHz
event rate (40 MHz interaction rate) by a factor of 200 with a maximum
delay of 1.2 $\mu$s. It works in three phases:

\begin{itemize}
  
\item {\bf Pretrigger:} Pretriggers originate from three sources:
  coincidences of pads and pixels in the 3rd and 4th superlayers of
  the muon detector, high-p$_t$ clusters in the calorimeter,
  coincidence patterns between the 3 pad chambers in the magnet. The
  three pretrigger systems produce ``messages'' which define a
  geometrical area (region-of-interest, or RoI) and an estimate of
  momentum. When operating at design rates, several such RoIs are
  expected per event.
  
\item {\bf Tracking:} Messages from the pretriggers are routed to a
  parallel pipelined network of custom-built processors which attempt
  to track them through 4 of the 6 main tracker superlayers behind the
  magnet (and, for muon pretriggers, through the superlayers of the
  muon system).  The processors map the superlayers geographically:
  each processor takes inputs from the 3 views of a contiguous region
  of a single superlayer.  Messages are passed from processor to
  processor.  In each processor, a search is made for hits inside the
  RoI and, when found, a new message is generated with refined track
  parameters and sent to the next processor (or processors when an RoI
  spans a boundary).
  
\item {\bf Decision:} Messages arriving at the furthest upstream
  superlayer are ``tracks'' with parameters determined with a typical
  accuracy of a single cell width in the outer tracker and 4 strip
  widths in the inner tracker. These messages are collected in a
  single processor where they are sorted by event (at any given time,
  100 events are being processed). A trigger decision is made based on
  the kinematics of single tracks and pairs of tracks. Events must be
  fully processed in less than 1.2 $\mu$s.

\end{itemize}

{\bf Status:} The inner and middle regions of the calorimeter are
equipped with pretrigger electronics. First operation began more than
a year ago and now, they are in routine operation.  Completion is
scheduled for end-January, 2000.  The muon pretriggers have been
tested with 15\% coverage.  Preliminary analysis indicates that the
system is performing up to expectations. Completion is scheduled for
mid-February. The high-p$_t$ system is under construction.
Installation and commissioning will begin in January.

Production of the track finding units is nearing completion and
installation is underway.  A slice has been installed and exercised,
analysis of data taken is underway.  Full system commissioning will
begin in January. If all goes well, the system should be at full power
by March, 2000.

The trigger decision unit has been in routine use since summer of this
year. Messages from the calorimeter pretriggers are routed directly
to the unit which sorts them by event and triggers when more than two
messages for one event are received. 

\subsection{Second and third level triggers, data acquisition}

The SLT \cite{hbhlt} is designed to work at  an input rate of 50\,kHz
and supply a suppression of at least 100 and up to 1000 for trigger
modes in which a detached secondary vertex is required.  The 3rd level
trigger is intended to provide a suppression of a factor of 10 on
trigger types for which RoI-based cuts to not provide sufficient
suppression -- e.g.  for events with no detached secondary vertex.

The SLT works on RoIs defined by the FLT, first gathering all hits
within the RoI from all tracking layers behind the magnet and
performing a fit. Successfully fit tracks are projected to and tracked
through the vertex detector.  At the end of the tracking process, a
vertex fit is performed on track pairs. Also, the impact parameters of
tracks relative to the target wires are estimated.  Trigger decisions
are made based on the outcome of the vertex fit and/or the track
impact parameters.

The 2nd/3rd level trigger and data acquisition are integrated into a
single system, implemented as a 240-node farm of standard Pentium
processors running the Linux operating system, a high bandwidth
switch, and a system of buffers (the ``second level buffer'') which
store event data while the second level trigger decision is being
made.  Both switch and buffers are built from the same DSP-based
board.

Upon acceptance by the first level, an event is transfered to the 2nd
level buffer and a processor node is assigned. The selected node
performs the 2nd level algorithm, requesting any needed data by
sending messages to the appropriate buffer module.  Data of events
passing 2nd level cuts are read from the 2nd level buffers into
processor memory and the 3rd level trigger is performed. Events
accepted at the third level are sent to the 4th level farm.

{\bf Status:} The second level farm has been operating routinely for
more than a year. The number of nodes in use currently stands at 80.
All 240 nodes are installed and cabled to the switch. The installation
will be finalized by end-January.

Portions of the 2nd level trigger algorithm have been exercised
routinely in the last year. Events triggered at the 1st level as
described in the previous section are transfered to the buffers.  The
assigned processor reads in the calorimeter data and searches for
clusters with p$_t$ above 1\,GeV. RoIs are generated and input into
the vertex tracking code which requests hits from selected regions of
the 2nd level buffers. Events for which at least one silicon track is
found are transfered to mass storage.

\begin{figure}[htb]
\begin{center}
\epsfig{file=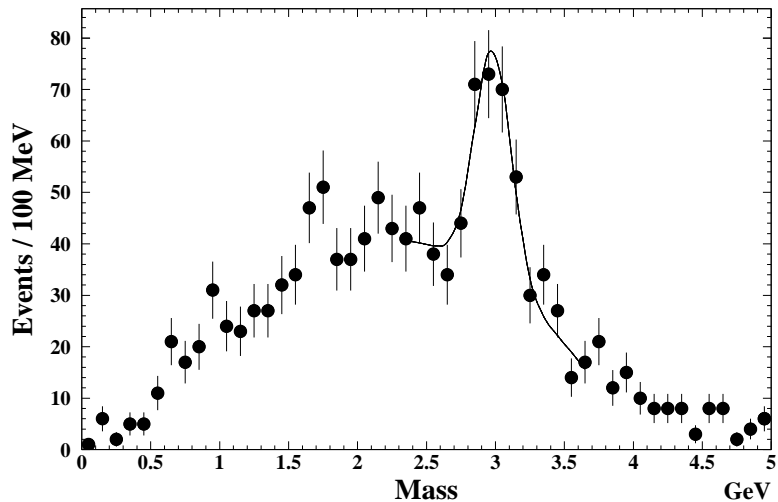,width=11cm}
\caption{Two track invariant mass}
\label{fig:jpsi}
\end{center}
\end{figure}

The accumulated data were then analyzed by the offline analysis group.
A plot of the mass of two silicon tracks (electron mass assumed) which
match high-p$_t$ calorimeter clusters and have an electron signature
in the calorimeter is shown in Fig.~\ref{fig:jpsi}. A clear J/$\psi$
peak is seen.

\subsection{4th level trigger}

The 4th level trigger is primarily intended for full online event
reconstruction. Like the 2nd level, it consists of a farm of Pentium
processors running Lynux. Unlike the 2nd level farm, it relies on
standard ethernet technology for communication and data transfer.  The
design input rate is 50 Hz.

{\bf Status:} In the last year, the event stream has been flowing
through the 4th level farm on its way to mass storage.  Several nodes
are in use for doing partial reconstruction and data monitoring. The
200 node farm is complete and in use, in part for monitoring tasks and
also for offline processing of data.  The full reconstruction code
exists and is being tuned in preparation for running on the completed
farm in January.

\section{Summary}

HERA-B has suffered setbacks resulting primarily from unexpected aging
effects in both inner and outer trackers. Solutions to these problems
have been found and the spectrometer is nearing completion. In the
meantime, considerable operational experience has been accrued in the
running periods between monthly 3-day shutdowns for installation.
After completion of the inner tracker in February, the detector will
be ready for the start of the HERA-B physics program. The critical
path is defined by the installation and commissioning of the first
level trigger. We look forward to the start of data-taking for physics
in February, 2000.

%%%%%%%%%%%%%%%%%%%%%%%%%%%%%%%%%%%%%%%%%%%%%%%%%%%%%%%%%%%%%%%%%%%%%%%%%
%%
%%   use this format to include a LaTeX table  into your paper
%%
%\begin{table}[htb]
%\begin{center}
%\begin{tabular}{l|llr}  
%A & B & C & D \\ 
%\end{tabular}
%\caption{An example of a small table.}
%\label{tab:ABCD}
%\end{center}
%\end{table}
%%%%%%%%%%%%%%%%%%%%%%%%%%%%%%%%%%%%%%%%%%%%%%%%%%%%%%%%%%%%%%%%%%%%%%%%%%%

\end{document}